\documentstyle[prl,aps,twocolumn]{revtex}

\begin{document}
\title{No-Concentrating Theorem of Pure Entangled States}
\author{Chuan-Wei Zhang\thanks{%
Electronic address: cwzhang@mail.ustc.edu.cn}, Chuan-Feng Li\thanks{%
Electronic address: cfli@ustc.edu.cn}, and Guang-Can Guo\thanks{%
Electronic address: gcguo@ustc.edu.cn}}
\address{Laboratory of Quantum Communication and Quantum Computation and Department\\
of Physics, \\
University of Science and Technology of China,\\
Hefei 230026, People's Republic of China\vspace*{0.3in}}
\maketitle

\begin{abstract}
\baselineskip14ptSuppose two distant observers Alice and Bob share a pure
biparticle entangled state secretly chosen from a set, it is shown that
Alice (Bob) can probabilistic concentrate the state to a maximally entangled
state by applying local operations and classical communication (LQCC) if and
only if the states in the set share the same marginal density operator for
her (his) subsystem. Applying this result, we present probabilistic
superdense coding and show that perfect purification of mixed state is
impossible using only LQCC on individual particles.

\end{abstract}

\baselineskip13.5pt The deep ways that quantum information differs from
classical information involve the properties, implications, and uses of {\it %
quantum entanglement }$[1]$. As a useful physical resource of quantum
information, entanglement plays a key role for quantum computation $[2]$,
quantum teleportation $[3]$, quantum superdense coding $[4]$ and certain
types of quantum cryptography $[5]$. The manipulation of entangled states,
that is, the transformations between different entanglements, may have
fundamental importance in quantum information theory. Attempts have been
made to uncover the fundamental laws of the manipulations under local
quantum operations and classical communication (LQCC) $\left[ 6-21\right] $.
A remarkable process involving such manipulations is {\it concentration of
entanglement} $\left[ 6-10\right] $. To function optimally, many
applications of entanglement $\left[ 2-5\right] $ require maximally
entangled states. Unfortunately interactions with the environment always
occur, and will degrade the quality of the entanglement. But the environment
does not always destroy entanglement completely. The resulting states may
still contain some residual entanglement. The task is then to concentrate
this residual entanglement with the aim of obtaining maximally entangled
states.

All previous entanglement manipulation protocols only deal with a known
finite-dimensional entangled state shared by distant observers Alice, Bob,
Clair, etc. Thus a natural question arises: may we manipulate a set of
entangled states only by same LQCC, just like quantum clone $[22-25]$? Then,
if can, what property characterizes the set of entangled states to be
transformed? In this letter, we will investigate the problem with the
example probabilistic entanglement concentration in bipartite system $\left[
6,9\right] $. It is shown that two pure bipartite entangled states shared by
distant observers Alice and Bob in Hilbert space $C^N\otimes C^N$ of a
composite system $AB$ can be probabilistic concentrated to the maximally
entangled states by the same LQCC if and only if they share the same
marginal density operator $\rho _A$ or $\rho _B$ for Alice's or Bob's
subsystem and the local operations must be performed on corresponding
subsystem. The result means that Alice (Bob) can not probabilistically
concentrate entangled states that are different in her (his) local
observation.

Suppose Alice and Bob share a pure bipartite entangled state secretly chosen
from a set. For a bipartite state $\left| \Psi \right\rangle _{AB}$ acting
on $C^N\otimes C^N$ of a composite system $AB$, its Schmidt decomposition
has the standard form $\left| \Psi \right\rangle _{AB}=\sum\limits_{i=1}^N%
\sqrt{\lambda _i}\left| i\right\rangle _A\left| i\right\rangle _B$, where $%
0\leq \lambda _i\leq 1$, $\sum_i\lambda _i=1$, and $\left| i\right\rangle _A$
$\left( \left| i\right\rangle _B\right) $ form an orthogonal basis for
system $A$ $\left( B\right) $. Here we denote $\lambda _i$ are ordered
decreasingly, i.e., $\lambda _1\geq \lambda _2\geq ...\geq \lambda _N$. Note
that all phases have been absorbed in the definition of the states $\left|
i\right\rangle _A$ so that the $\lambda _i$ are positive real numbers.
Furthermore a general pure bipartite entangled state can be represented as $%
\left| \varphi \right\rangle _{AB}=U_A\otimes U_B\left| \Psi \right\rangle
_{AB}$, where $U_A\ $and $U_B$ are local unitary transformations by Alice
and Bob respectively. Obviously $\left| \varphi \right\rangle _{AB}$ and $%
\left| \Psi \right\rangle _{AB}$ share same Schmidt decomposition
coefficients. The marginal density operators for Alice's and Bob's
subsystems are defined respectively as $\rho _A=tr_B\left| \varphi
\right\rangle \left\langle \varphi \right| $ and $\rho _B=tr_A\left| \varphi
\right\rangle \left\langle \varphi \right| $. The standard $N$-dimensional
maximally entangled state can be denoted as $\left| \Phi _N\right\rangle
_{AB}=\frac 1{\sqrt{N}}\sum\limits_{i=1}^N\left| i\right\rangle _A\left|
i\right\rangle _B$. Similarly all the states $U_A\otimes U_B\left| \Phi
_N\right\rangle _{AB}$ are also maximally entangled states.

{\bf Theorem 1:}{\sl \ }Two different pure entangled states can be
probabilistic concentrated to the maximally entangled state by the same LQCC
if and only if they share same marginal density operators $\rho _A$ or $\rho
_B$ for Alice's or Bob's subsystem.

{\sl Proof: }Generally, the two states to be concentrated can be represented
as $\left| \Psi \right\rangle _{AB}=\sum\limits_{i=1}^N\sqrt{\lambda _i}%
\left| i\right\rangle _A\left| i\right\rangle _B$ and $\left( U_A\otimes
U_B\right) \left| \Omega \right\rangle _{AB}=\left( U_A\otimes U_B\right)
\sum\limits_{i=1}^N\sqrt{\mu _i}\left| i\right\rangle _A\left|
i\right\rangle _B$. The most general scheme of entanglement manipulations of
a bipartite pure entangled state involves local operations of respective
system and two-way communications between Alice and Bob. The local
operations can be represented as generalized measurements, described by
operators $A_k$ and $B_l$ on each system, satisfying the condition $%
\sum_kA_k^{+}A_k\leq I_N$ and $\sum_lB_l^{+}B_l\leq I_N$, where $I_N$ is the
unit operator of Alice or Bob subsystem. The LQCC protocols we consider map
the initial state $\left| \varphi \right\rangle _{AB}\left\langle \varphi
\right| $ to the maximally entangled state 
\begin{equation}
\left| \Phi _N^{^{\prime }}\right\rangle _{AB}\left\langle \Phi _N^{^{\prime
}}\right| =\frac{\sum_{kl}A_k\otimes B_l\left| \varphi \right\rangle
_{AB}\left\langle \varphi \right| A_k^{+}\otimes B_l^{+}}{Tr\left(
\sum_{kl}A_k\otimes B_l\left| \varphi \right\rangle _{AB}\left\langle
\varphi \right| A_k^{+}\otimes B_l^{+}\right) }.  \eqnum{1}
\end{equation}
The initial and final states are pure, it follows that 
\begin{equation}
A_k\otimes B_l\left| \varphi \right\rangle _{AB}=\sqrt{p_{kl}}\left| \Phi
_N^{^{\prime }}\right\rangle _{AB},  \eqnum{2}
\end{equation}
with non-negative probability $p_{kl}$ satisfying $p_{kl}=Tr\left(
A_k\otimes B_l\left| \varphi \right\rangle _{AB}\left\langle \varphi \right|
A_k^{+}\otimes B_l^{+}\right) $.

We first prove {\bf Theorem 1 }with the assumption that only Alice execute
the generalized measurement. Any operation in quantum mechanics can be
represented by a unitary-reduction evolution $U$ together with a
measurement. We demand the output states of the concentrating machine are
pure maximally entangled states. This requires the measurement should be
performed postselectively. We introduce a probe\ $P$\ in a\ $n_P$%
-dimensional Hilbert space\ ($n_P\geq 2$), and denote orthonormal states of
the probe\ $P$ as\ $\left| P_i\right\rangle $. The concentrations may be
successful for several $\left| P_i\right\rangle $ and the output states
should always be the maximally entangled state for each $\left|
P_i\right\rangle $, although it may not be the standard form. Any unitary
operator $V$ performed on Bob's subsystem to the maximally entangled state
is equivalent to the transpose $V^{+}$ by Alice. So after the measurement
Alice can always transfer the output states of system $AB$ into the standard
form and we need only consider the following evolution: 
\begin{equation}
\left( U_{AP}\otimes I_B\right) \left| \Psi \right\rangle _{AB}\left|
P_0\right\rangle =\sqrt{\gamma }\left| \Phi _N\right\rangle _{AB}\left|
P_1\right\rangle +\sqrt{1-\gamma }\left| \omega \right\rangle _{AB}\left|
P_0\right\rangle .  \eqnum{3}
\end{equation}

We measure the probe\ $P$\ after the evolution. The concentrating attempt
succeeds if and only if the measurement output of the probe is\ $P_1$. With
probability\ $\gamma $\ of success, this measurement projects the composite
system\ $AB$\ into the maximally entangled state $\left| \Phi
_N\right\rangle _{AB}$. The parameters\ $\gamma $\ is called the
concentrating efficiency. Lo and Sopescu $[9]$ have shown the maximum
probability $\gamma _{\max }=N\lambda _N$. Thus if we demand the
concentrating probability is no-zero, the minimum of Schmidt decomposition
coefficients $\lambda _N$ should be greater than zero. Our task remains to
search the sufficient and necessary conditions for that the state $\left(
U_A\otimes U_B\right) \left| \Omega \right\rangle _{AB}$ can also be
probabilistic concentrated by operator $\left( U_{AP}\otimes I_B\right) $
with the same postselective measurement. Obviously $\left( I_A\otimes
U_B\right) $ does not influence the marginal density operator for Alice's
subsystem that yields the interchange 
\begin{eqnarray}
&&\left( U_{AP}\otimes I_B\right) \left( I_A\otimes U_B\right) \left| \Omega
\right\rangle _{AB}  \eqnum{4} \\
&=&\left( I_A\otimes U_B\right) \left( U_{AP}\otimes I_B\right) \left|
\Omega \right\rangle _{AB}.  \nonumber
\end{eqnarray}
The interchange above does not violate the unitarity of the operator since
if we use the state $_{AB}\left\langle \Psi \right| \left( U_{AP}^{+}\otimes
I_B\right) $ to make inner-product with both sides of Eq. (4), the equation
still preserve. So we need only consider the state $\left( U_A\otimes
I_B\right) \left| \Omega \right\rangle _{AB}$. We first introduce two
unitary operators of Bob's side which help to describe the property of
operator $U_{AP}\otimes I$. Define operators $T_k$ and $S^i$ which act as
follows: 
\begin{eqnarray*}
T_k\left| j\right\rangle  &=&\left| \left( j+k\right) 
\mathop{\rm mod}
d\right\rangle , \\
S^i\left| j\right\rangle  &=&\left( -1\right) ^{\delta _{ij}}\left|
j\right\rangle .
\end{eqnarray*}
\ Together with Eq. (3) and (4), the linearity of the operators yields 
\begin{eqnarray}
&&\ \ \ \ \ \ \ \left( U_{AP}\otimes I\right) \sqrt{4\lambda _k}\left|
k\right\rangle _A\left| n\right\rangle _B\left| P_0\right\rangle   \eqnum{5}
\\
\  &=&\left( U_{AP}\otimes I\right) \left( I\otimes \left( I-S^n\right)
T_{\left( n-k\right) 
\mathop{\rm mod}
d}\right) \left| \Psi \right\rangle _{AB}\left| P_0\right\rangle   \nonumber
\\
\  &\rightarrow &\sqrt{\gamma }\left( I\otimes \left( I-S^n\right) T_{\left(
n-k\right) 
\mathop{\rm mod}
d}\right) \left| \Phi _N\right\rangle _{AB}\left| P_1\right\rangle  
\nonumber \\
\  &=&\sqrt{4\gamma /N}\left| k\right\rangle _A\left| n\right\rangle
_B\left| P_1\right\rangle .  \nonumber
\end{eqnarray}
Supposing $U_A=\sum\limits_{i,j}a_{ij}\left| i\right\rangle \left\langle
j\right| $, we derive the evolution equation of state $\left( U_A\otimes
I_B\right) \left| \Omega \right\rangle _{AB}$ on the unitary operation $%
\left( U_{AP}\otimes I_B\right) $ as 
\begin{eqnarray}
&&\ \ \ \ \ \ \ \left( U_{AP}\otimes I\right) \left( U_A\otimes I\right)
\left| \Omega \right\rangle _{AB}\left| P_0\right\rangle   \eqnum{6} \\
\  &=&\left( U_{AP}\otimes I\right) \left( \sum_{k=1}^N\sqrt{\mu _k}\left(
\sum_{i=1}^Na_{ik}\left| i\right\rangle _A\left| k\right\rangle _B\right)
\right) \left| P_0\right\rangle   \nonumber \\
\  &\rightarrow &\sqrt{\frac \gamma N}\sum_{k=1}^N\sum_{i=1}^Na_{ik}\sqrt{%
\frac{\mu _k}{\lambda _i}}\left| i\right\rangle _A\left| k\right\rangle
_B\left| P_1\right\rangle .  \nonumber
\end{eqnarray}

If the local unitary operator $U_{AP}\otimes I_B$ can concentrate both the
states $\left| \Psi \right\rangle _{AB}$ and $\left( U_A\otimes I_B\right)
\left| \Omega \right\rangle _{AB}$, the final state of system $AB$ in Eq.
(6) should be the maximally state, which means that 
\begin{equation}
\sum_{i=1}^Na_{ik}^{*}a_{il}\sqrt{\frac{\mu _k}{\lambda _i}}\sqrt{\frac{\mu
_l}{\lambda _i}}=C\delta _{kl}\text{ }  \eqnum{7}
\end{equation}
where $C$ is a constant. Denote matrix $\mu =diag\left( \mu _1,\mu
_2,...,\mu _N\right) $, $\lambda =diag\left( \lambda _1,\lambda
_2,...,\lambda _N\right) $, Eq. (7) is equivalent to the following matrix
equation 
\begin{equation}
C\lambda =U_A\mu U_A^{+}  \eqnum{8}
\end{equation}
Since the eigenvalues of matrix $\mu $ are invariant on the unitary
transformation $U_A$ and $\sum_i\lambda _i=1$, $\sum_i\mu _i=1$, Eq. (8) can
be rewritten as 
\begin{equation}
\lambda =U_A\mu U_A^{+}=\mu ,  \eqnum{9}
\end{equation}
Consequently, 
\begin{equation}
\rho _A=\sum_i\lambda _i\left| i\right\rangle \left\langle i\right|
=U_A\left( \sum_i\mu _i\left| i\right\rangle \left\langle i\right| \right)
U_A^{+}=\rho _A^{^{\prime }}  \eqnum{10}
\end{equation}
where $\rho _A=tr_B\left| \Psi \right\rangle \left\langle \Psi \right| $ and 
$\rho _A^{^{\prime }}=tr_B\left( U_A\otimes I\right) \left| \Omega
\right\rangle \left\langle \Omega \right| \left( U_A^{+}\otimes I\right) $
are the marginal density operators for Alice's subsystem. So we have proven
the necessary condition. Since $U_A\mu U_A^{+}=\mu $, it can be easily
proven that the operation $U_A$ is equivalent to the transposed operation $%
U_A^{+}$ done by Bob. Because of the interchange of Eq. (4), Eq. (6) can be
realized in physical means, so we prove the sufficient condition.

Now the remained problem needed to be considered is whether the generalized
measurement in Bob side could make it possible to concentrate both the two
states. Consider the following facts\footnote{%
Given any pure biparticle state $\left| \Psi \right\rangle _{AB}$ shared by
Alice and Bob and any complete set of projection operators $\left\{
P_l^{Bob}\right\} $'s by Bob, there exists a complete set of projection
operators $\left\{ P_l^{Alice}\right\} $'s by Alice and, for each outcome $l$%
, a direct product of local unitary transformations $U_l^A\otimes U_l^B$
such that, for each $l$%
\[
\left( I\otimes P_l^{Bob}\right) \left| \Psi \right\rangle _{AB}=\left(
U_l^A\otimes U_l^B\right) \left( P_l^{Alice}\otimes I\right) \left| \Psi
\right\rangle _{AB} 
\]
}
\begin{equation}
A_k\otimes B_l\left| \Psi \right\rangle _{AB}=\left( A_kV_l^AB_l\otimes
V_l^B\right) \left| \Psi \right\rangle _{AB},  \eqnum{11}
\end{equation}
\begin{eqnarray*}
&&\left( A_k\otimes B_l\right) \left( U_A\otimes U_B\right) \left| \Omega
\right\rangle _{AB} \\
&=&\left( A_kU_AH_l^AB_lU_B\otimes H_l^B\right) \left| \Omega \right\rangle
_{AB},
\end{eqnarray*}
where $V_l^A$, $V_l^B$, $H_l^A$ and $H_l^B$ are local unitary operations.
Above two equations means that $A_k$ can concentrate both states $\left(
V_l^AB_l\otimes I\right) \left| \Psi \right\rangle _{AB}$ and $\left(
U_AH_l^AB_lU_B\otimes I\right) \left| \Omega \right\rangle _{AB}$ to the
maximally entanglement states. From above proof we know states $\left(
B_l\otimes I\right) \left| \Psi \right\rangle _{AB}$ and $\left(
B_lU_B\otimes I\right) \left| \Omega \right\rangle _{AB}$ must have same
Schmidt decomposition coefficients. Generally we could write 
\begin{eqnarray}
\left( B_l\otimes I\right) \left| \Psi \right\rangle _{AB} &=&\sqrt{%
\varepsilon }\left( E_1\otimes F_1\right) \sum_{i=1}^N\sqrt{\kappa _i}\left|
i\right\rangle \left| i\right\rangle ,  \eqnum{12} \\
\left( B_lU_B\otimes I\right) \left| \Omega \right\rangle _{AB} &=&\sqrt{%
\epsilon }\left( E_2\otimes F_2\right) \sum_{i=1}^N\sqrt{\kappa _i}\left|
i\right\rangle \left| i\right\rangle ,  \nonumber
\end{eqnarray}
where $E_i\otimes F_i$ are local unitary operations, $\varepsilon $ and $%
\epsilon $ are the probabilities of success. Above two equations could be
represented using matrixes as 
\begin{eqnarray}
B_l\sqrt{\lambda } &=&\sqrt{\varepsilon }E_1\sqrt{\kappa }F_1^{+}, 
\eqnum{13} \\
B_lU_B\sqrt{\mu } &=&\sqrt{\epsilon }E_2\sqrt{\kappa }F_2^{+},  \nonumber
\end{eqnarray}
where $\kappa =%
\mathop{\rm diag}
\left( \kappa _1,\kappa _2,...,\kappa _N\right) $. Thus we obtain 
\begin{equation}
T^{+}\kappa T=F_2\kappa F_2^{+},  \eqnum{14}
\end{equation}
where $T=\sqrt{\frac \varepsilon \epsilon }F_1^{+}\sqrt{\lambda ^{-1}}U_B%
\sqrt{\mu }E_2^{+}F_2$. Eq. (14) means that $T$ is unitary, it follows 
\begin{equation}
\frac \varepsilon \epsilon U_B\mu U_B^{+}=\lambda .  \eqnum{15}
\end{equation}
Since $\sum_i\lambda _i=1$, $\sum_i\mu _i=1$, we get $\varepsilon =\epsilon $%
, $\mu =\lambda $, $\rho _B=\rho _B^{^{\prime }}$ and complete the proof.

So far we have proven {\bf Theorem 1}, which gives a necessary and
sufficient condition to determine whether two states can be probabilistic
concentrated or not by same local actions and classical communication. For
general situation the theorem indicate that the ordered Schmidt coefficients
of the states to be concentrated must be same. But the two states need not
to be same, there can exist unitary transformations on both Alice's and
Bob's sides. While arbitrary on Bob's (Alice's) side, the unitary operator
on Alice's (Bob's) side must preserve the density matrix $\rho _A$ ($\rho _B$%
). That means only when there exist some coefficients satisfying $\lambda
_i=\lambda _{i+1}$, the unitary operator $U_A$ ($U_B$) can be no-unit. 

In the following we will apply above consequence to discuss some problems.
With the proof above it is obvious that quantum superdense coding can be
generalized to the probabilistic situation. In Eq. (4) we suppose Bob has
four choices to perform $U_B$ i.e. $\left\{ I,\sigma _x,i\sigma _y,\sigma
_z\right\} $, just like that in the original paper $[4]$. Bob still send his
particle to Alice after he has performed $U_B$. Alice's task is then to
identify the four states, whose optimal probability had been obtained by
Duan and Guo $[25]$.

In all there are two fundamentally different types of concentration
protocols: collectively and individually: those acting on individual pairs
of entangled particles and those acting collectively on many pairs. In the
proof of {\bf Theorem 1} we also showed the following important result:

{\bf Proposition}: Two different entangled states cannot be transferred into
the same maximally entangled state by the same LQCC on individual pairs.

While one can always, with finite probability, bring an individual entangled
pure state to a maximally entangled state using only local operation, we show

{\bf Theorem 2 }: It is impossible to purify the mixed state to the
maximally entangled state by LQCC on individual pairs. That also means
perfect probabilistic teleportation using mixed state is impossible.

{\sl Proof: }The proof is simple. Now consider a given mixed state $\rho $,
generally we can use the spectral decomposition of the state $\rho
=\sum_ip_i\left| \psi _i\right\rangle \left\langle \psi _i\right| $. {\bf %
Proposition }indicates that the different decomposition term $\left| \psi
_i\right\rangle $ of mixed state $\rho $ can never be transferred into same
pure states, which means $\rho $ cannot be concentrated into a pure
entangled state by LQCC on individual pairs. In fact Linden {\it et al.} $%
[10]$ have shown that it is impossible to purify singlets, or even increase
the fidelity of a Werner density matrix infinitesimally, by any combination
of local actions and classical communication acting on individual pairs. Our
proof is more direct and general for high dimensional situation. This result
is surprising because we expect entanglement to be a property of each pair
individually rather than a global property of many pairs.

In summary, we have shown the pure bipartite entangled states secretly
chosen from a set in Hilbert space $C^N\otimes C^N$ can be probabilistic
concentrated to the maximally entangled states by the same LQCC if and only
if they share same marginal density operator for one of the two parties. The
physical meaning of this consequence is that both Alice and Bob cannot
probabilistic concentrate two states which are different to her (his)
observation. Using this result we proposed the probabilistic superdense
coding and showed that perfect purification of mixed state is impossible
using only LQCC on individual particles.

This work was supported by National Natural Science Foundation of China.

\end{document}